\documentclass[12pt]{article}

\usepackage{epsf}

\usepackage{epsfig}

\usepackage{wrapfig}

\usepackage{amsmath}

\usepackage{amssymb}

\usepackage{eufrak}

\begin{document}

%\begin{flushright}

% 539.12: 539.171 \end{flushright}

%\vskip 15mm

\title{Many-worlds interpretation of quantum theory and mesoscopic anthropic principle }

\author{A.Yu. Kamenshchik$^{1,2}$ and O.V. Teryaev$^{3}$}
\date{}
%\underline{C.D. Author2}$^{1,2}$ and

%V.A. Author3$^{1,3}$

\vskip 5mm
\maketitle
\hspace{-6mm}$^{1}$Dipartimento di Fisica and INFN, Via 
Irnerio 46, 40126 Bologna, Italy\\
$^{2}$L.D. Landau Institute for Theoretical Physics of the 
Russian Academy of Sciences, Kosygin str. 2, 
119334 Moscow, Russia\\
$^{3}$
Joint Institute for Nuclear Research, 141980 Dubna, Russia

\vskip 5mm

\begin{abstract}
We suggest to combine the Anthropic Principle with Many-Worlds 
Interpretation of Quantum Theory. Realizing the multiplicity of worlds
it provides an opportunity of explanation of some important events which are assumed to be extremely improbable. 
The Mesoscopic Anthropic Principle suggested here is aimed to explain appearance of such events which are necessary for emergence of Life and Mind. 
It is complementary to Cosmological Anthropic Principle explaining the fine tuning of fundamental constants.  
We briefly discuss various possible applications of Mesoscopic Anthropic Principle including the Solar Eclipses 
and assembling of complex molecules.  Besides, we address the problem of Time's Arrow in the framework of Many-World Interpretation.  We suggest the recipe for disentangling of  quantities defined by fundamental physical laws  and by an anthropic selection.

\end{abstract}

%\vskip 8mm

\section{Introduction}

The anthropic principle (AP) was proposed long ago \cite{Dirac,Dicke,Carter,Rosental,BT}
but recently it got a strong boost (see e.g. \cite{Rubakov,Weinberg})
connected with the development of cosmology \cite{cosmol}
and string theory \cite{string}.
The general idea of AP consists in the statement that existence of the (human) observer
imposes  important restrictions on the basic laws and fundamental physical constants.
As soon as these restrictions happen to be of tantamaunt importance, the required
values of physical constants appear to be extremely improbable.
This smallness of probability could be  compensated by the huge number of
universes constituting Multiverse. Under this term one should understand a complicated object which  may be formed by the process of the ramification of the spatial
structure of the universe due to the effects of spontaneous symmetry breaking producing inflationary expansion of the patches of spacetime.
Such an opportunity is inherent in the chaotic inflation models
\cite{Linde}.

Another source of multiversity is the existence of the so called string landscape which means that
the fundamental superstring theory contains  a huge amount of vacuum states, each of those
may lead to quite different universes with different physics.

Here we would like to discuss yet another source of
multiplicity opening the possibility of further extension of
applicability of AP. It corresponds to many-worlds interpretation  of quantum theory \cite{mw}.
As soon as this multiplicity does not lead to the change of fundamental cosntants we
are dealing with what we call "Mesoscopic" AP, corresponding to the scales intermediate between
cosmological and microscopic ones.

The structure of the paper is the following:
The second section is devoted to a brief review of the basic ideas of the many-worlds interpretation of quantum mechanics; in the third section we discuss branching of worlds understood in the sense of the
defactorization of the wave function and the problem of the preferred basis; in the fourth section
we consider the important  problem of irreversibility and appearance of the arrow of time in terms of
the many-worlds interpretation; the fifth sections deals with the definition of the Mesoscopic
Anthropic Principle and its simplest applications to planetary systems; in the sixth section we treat
biological evolution in terms of variety of options provided by the quantum evolution; in the last section
we discuss the main results and suggest some criteria for disentangling of  quantities defined by fundamental physical laws  and by an anthropic selection.

\section{Many-worlds interpretation of quantum mechanics}

The many-worlds interpretation (MWI) of quantum mechanics  was suggested by H. Everett in 1957 \cite{Everett}
and its invention was motivated by two factors. One of them was intensively discussed since the moment
of creation of quantum mechanics: it is the problem of reconciliation between two processes
present in the theory - dynamical evolution in  accordance with the Schr\"odinger equation and the
reduction of wave packet, responsible for an observation of the unique outcome of quantum measurement
when the quantum state represents a superposition of the corresponding eigenstates.
In the most popular Copenhagen interpretation of quantum mechanics such a coexistence of these
two processes was provided by the separation of the so called classical realm, which in some versions
was connected even with the presence of conscious observer. Thus the desire of getting rid of the ambiguity
connected with the wave packet reduction postulate and having a unique quantum description of Nature
stimulated the creation of MWI. In the framework of MWI the 
Schr\"odinger evolution is the only process,
 the principle of superposition is applicable to all the states including macroscopic ones and all the
outcomes of any measurement-like processes are realized simultaneously but in different ``parallel universes''. The very essence of the many-worlds interpretation can be expressed by one simple formula we are about to derive.
Let us consider the wave function of a system, containing two subsystems (say, an object and a device),
whose wave functions are respectively $|\Phi\rangle$ and $\Psi\rangle$ and let us the process of ithe interaction between these two subsystems is described by a unitary operator $\hat{U}$. The result of action of this operator can be represented as
\begin{equation}
\hat{U}|\Phi\rangle_0 \Psi\rangle_{i} = |\Phi\rangle_i \Psi\rangle_{i}.
\label{unitary}
\end{equation}
Here the state $|\Psi\rangle_{i}$ is a quantum state of the object corresponding to a definite outcome of the
experiment, while $|\Phi\rangle_0$ is an initial state of the measuring device. Now, let the initial state
of the object be described by a superposition of quantum states:
\begin{equation}
|\Psi\rangle = \sum_{i} c_i |\Psi\rangle_i.
\label{superposition}
\end{equation}
That superposition principle immediately leads to
\begin{equation}
\hat{U}|\Phi\rangle_0 \Psi\rangle = \hat{U}|\Phi\rangle_0\sum_{i} c_i |\Psi\rangle_i =
\sum_{i}c_i|\Phi\rangle_i \Psi\rangle_i.
\label{superposition1}
\end{equation}
Here $|\Phi\rangle_i$ describes the state of the measuring device, which has found the quantum object in the
state $|\Psi\rangle_i$. The superposition  (\ref{superposition1}) contains more than one term, while one
sees only one outcome of measurement. The reduction of the wave packet postulate solves this puzzle by
introducing another process eliminating in a non-deterministic way all the terms in the right-hand side
of Eq. (\ref{superposition1}) but one. The MWI instead says that all the terms  of the superposition
are realized but in different universes.

The MWI looks the most consistent between interpretations of quantum theory, because it
ultimately reduces the number of postulates. Moreover, one of the proponents of MWI B.S. DeWitt says that in the framework of it
the mathematical formalism of the theory gives itself its interpretation \cite{DeWitt}.

Now, let us turn to second motivation for MWI. In quantum cosmology there is no external observer and
hence, no, classical realm. Thus, MWI  matches quite well the quantum cosmology.

The many-worlds interpretation with its branching of universes apparently opens a magnificient
opportunities for the application of the AP. This possibility was practically overlooked in the literature
(see, however \cite{BKP}).

\section{Branching of Worlds and  the  preferred basis}

The opportunity to extract non-trivial physical consequences in the context of MWI is based on the
treating of the branching of worlds as an objective process. However, inevitable question arises:
decomposing the wave function of the universe one should choose a certain basis. The result of
the decomposition essentially depends on it. Thus, the so called  problem of the choice of the preferred basis arises \cite{basis} The essence of the problem can be easily formulated considering
the same example of a quantum system consisting of two subsystems. Let us emphasize that now we would like
to undertake a consideration of a general case without particular reference to artificial measuring devices
and quantum objects (for a moment we consider this division of a system into subsystems as granted).
The only essential characteristics of the branching process is the defactorization of the wave function.
That means that if at the initial moment the wave function of the system under consideration was represented
by the direct product of the wave functions of the subsystems
\begin{equation}
|\Psi\rangle = |\phi\rangle |\chi\rangle
\label{direct}
\end{equation}
then after an interaction between the subsystems it becomes
\begin{equation}
\sum_{i} c_i |\phi\rangle_i |\chi\rangle_i,
\label{defact}
\end{equation}
where more than one coefficient $c_i$ is differentt from zero.
Apparently the decomposition (\ref{defact}) can be done in various manners. As soon as each term
is associated with a separate universe, the unique prescription for the construction of such a
superposition should be fixed.  We believe that the correct choice of the preferred basis is the  so
called Schmidt or bi-orthogonal basis. This basis is formed by eigenvectors of both the density matrices of the subsystems of the quantum system under consideration.
These density matrices are defined as
\begin{equation}
\hat{\rho}_{I} = Tr_{II}|\Psi\rangle \langle \Psi|,
\label{density}
\end{equation}
\begin{equation}
\hat{\rho}_{II} = Tr_{I}|\Psi\rangle \langle \Psi|.
\label{density1}
\end{equation}
Remarkably, the eigenvalues of the density matrices coincide and hence the number of non-zero eigenvalues is the same, in spite of the fact that the corresponding Hilbert spaces can be very different.
\begin{equation}
\hat{\rho}_{I} |\phi_n\rangle = \lambda_n |\phi_n\rangle,
\label{density2}
\end{equation}
\begin{equation}
\hat{\rho}_{II} |\chi_n\rangle = \lambda_n |\chi_n\rangle,
\label{density3}
\end{equation}
Consequently, the wave function is decomposed as
\begin{equation}
|\Psi\rangle = \sum{\alpha} \sqrt{\lambda_{n}}|\phi_n\rangle|\chi_n\rangle.
\label{density4}
\end{equation}

The bi-orthogonal basis
was first used at the dawn of quantum mechanics by E. Schr\"odinger \cite{Schrod}  for study of correlations between
quantum systems and was applied to MWI in \cite{Zeh,we}.
Recently, this basis is actively used for measuring of degree of entanglement, in particular, in relation
to quantum computing \cite{Schmidt-comp}. The expansion with respect to eigenvectors of spin density matrix
and density matrix positivity was also used in hadronic physics and non-perturbative QCD \cite{Teryaev,Teryaev1}.

We believe that the bi-orthogonal basis being defined by the fixing of the decomposition of the system
into subsystems have a fundamental character and determines the worlds which result from the defactorization process. However, the subdivision of the system onto subsystems which implies the branching of the worlds
should satisfy some reasonable criteria which we are not ready to formalize at the moment (see, however \cite{kam} for analysis of some relatively simple cases). One can say, that the decomposition into the subsystems should be such that the corresponding preferred basis were rather stable. For example, when one treats a quantum mechanical expreriment of the Stern-Gerlach type, it is natural to consider the measuring device and the atom as subsystems.

\section{Time`s arrow}
The formalism of the many-world interpretation of quantum theory permits to reformulate the problem of a direction of time in a very transparent way. Indeed, the basic dynamics equations are invariant with respect to the operation of time reflection, while the macroscopic phenomena shows the irreversibility or the presence of the arrow of time.  One of the quantitative manifestations of these phenomena is the growth of the von Neumann entropy \cite{von}
\begin{equation}
S = -Tr (\hat{\rho}\ln\hat{\rho}) = -\sum_i \lambda_i \ln \lambda_i \equiv \sum_i S_i.
\label{entropy}
\end{equation}
where the last equality introduces, in the context of MWI  the notion of relative entropies of branches.
This entropy is minimal and equal to zero for a pure quantum state. Usually, the presence of the arrow of time is connected with the existence of some additional constraints on the solutions of fundamental equations. For example, choosing an initial state as a state with low value of entropy, one naturally sees
its growth.
We make an observation that the branching process in the MWI naturally produces the states with a smaller initial  relative entropy (that is calculated by taking into account only one branch). In other words, after the measurement-like act of branching a new branch is in factorized quantum state and the density matrices of
all its subsystems correspond to pure quantum states. This does not contradict to the increase of entropy
in the standard (Copenhagen) treatment of quantum measurement. In the latter case one is dealing after
the measurement with the classical statistical mixture of a various outcomes producing increase of entropy
which can be measured experimentally. At the same time in MWI the process of measurement (defactorization of the wave function) naturally implies the inclrease of entropy, but after the identification of an outcome of measurement, when the defactorization of the wave function is completed, the relative entropy
(related to the branch where we live) becomes equal to $S_i$. Forgeting about other branches, which is equivalent to the reduction of wave packet in the Copenhagen interpretation, corresponds to rescaling $\lambda \to 1$
and $S_i \to S^R; S^R(t_0)=0$, where $S_R$ is the redefined entropy after the branching happened at
time $t_0$
Thus, relative entropy of each branch
is always growing, $S_i^R(t)> S_i^R(0)=0$,  so is $S_i$ and the usual measurable entropy of classical statistical mixture
which is just the sum (\ref{entropy}) of the entropies of the branches. Note that this nullification of relative entropy does not involve the distant regions of Universe which are the same for all the branches.

Thus, MWI provides another manifestation of the effect of boundary conditions which is present
in any explanation of irreversibility. The example of such boundary conditions
is, say, the correlations weakening in the BBGKI chain of equations leading to the
appearance of irreversibility. In another approach, when deriving \cite{zasl}
the irreversible master equation
from the reversible Kolmogorov-Chapman equation is is sufficient\cite{Teryaev2} to assume the existence
of the initial conditions in the past.
The role of boundary effects for the
irreversibility  of field theory evolution equations implying the "scale arrow" , analogous to time's arrow, is discussed in \cite{Teryaev2,Teryaev1}. In turn, the irreversibility with respect to time reflection
in field theory may appear
either because of T(or CP) violation at the fundamental level or because of its
simulation by imaginary phases of scattering amplitudes. The latter crucially depend
on the sign of $i \epsilon$ in the Feynman propagators which is imposed by the causal boundary conditions for Green functions. This effect is giving rise to T-odd spin asymmetries \cite{todd} being the subject of intensive theoretical and experimental studies.

In the actual case of MWI the choice of boundary conditions corresponds to the choice of factorized
wave function in the past, rather than in the future.
However, as MWI may be considered as "self-interpretation" of the mathematical formalism of quantum theory \cite{DeWitt}, the suggested approach may explain the fundamental phenomenon of Arrrow of Time in a similar manner.

\section{Planetary Coincidences and Mesoscopic Anthropic Principle}

It is usually believed that the suitable values of fundamental constants are sufficient
for emergence of stars, planetary systems and all the astrophysical objects required for
apperance of life. However, there are a number of observations pointing to the special, privileged, role
of the Solar system (see e.g. \cite{planet}). All the values
describing this privileged position
cannot involve the fine-tuning of neither constants of elementary particle physics nor cosmology.
Therefore we call such a coincidences the mesoscopic anthropic coincidences and
the related selection the Mesoscopic Anthropic Principle (MAP).

The first natural opportunity to find the priviliged values of planetary characteristics
is to explore the vast number of galaxies stars, and planets in our Universe \cite{muller}.
Note, that the necessity of this large number provides a sort of answer for one line
of possible criticism of AP suggesting that the existence of such a
large Universe is hardly necessary for the life on the Earth, this argument being
best expressed by S. Hawking who was saying that �''our Solar system is certainly a prerequisite for our existence, But there does not seem any necessity for other galaxies to exist''.

At the same time, the selection among the large number of distant astrophysical objects
does not seem sufficient if some fine-tuned value of mesoscopic parameter is required.
For this aim the small changes of the relevant parameter within the required range
are important. This is exactly what happens in the chaotic inflation or stringy landscape
and allows for a fine tuning of fundamental constants
\footnote{Such a small changes of some parameter constitute, in fact, the cornerstone of Darwinian natural selection, see also the next section}.

As a possible solution of this problem we suggest the MWI is a source of
small variations of mesoscopic planetary constants in different worlds.
We assume that the measurement-like quantum interactions leading to the branching
occur all the time independently of the presence of (conscious) observer and produce
the planetary systems in parallel Everett worlds whose parameters differ by small amount.

The example of planetary fine-tuning is provided by Solar eclipses requiring the
coincidences of angular sizes of Sun and Moon, as seen from the Earth.
There is currently no explanation of this coincidence, apart from teleological arguments
\cite{planet}. At the same time, this coincidence would be explained if the
eclipse were necessary for some stage of the emergence of life. This
does not seem completely impossible, although there is no evidences in favour of such
a relation. One possibility is the emergence of life due to photochemical reaction requiring
the shadowing of strong ultraviolet radiation of the Sun but presence of the radiation of Solar
Corona. Should such or similar scenario find the experimental support (which is possible, at least in principle) this would mean also the support of MAP and the role of MWI.

\section{Mesoscopic Anthropic Principle and Biological Evolution}

In turn, even suitable planetary environment and emergence of primitive life does not,
contrary to popular wisdom, leads to the appearance of its complex forms.
The Darwinian evolution is an adaptive one \cite{dawkins} and explains the
arising of the complex structures if they provide the evolutionary advantages.
At the same time, the appearance of complex structures, which does not lead to immediate evolutionary success,
including the Human beings
is not trivial to explain. The production of complexity in the process of the type of
random walk may be explained \cite{gould} only if this complexity is relatively low. The random walk in that case is limited by zero complexity barrier and produces its increase. The further
evolutionary process explains the progress of most numerous species, like insects,
but not the apperance of  complex and rare ones. Therefore, the origin of humans, being the most
popular success of originals  theory of Darwin and Wallace, remains out of scope of its modern version.

The natural way to explain the appearance of very complex and improbable structures is provided by MWI.
This opportunity was recently explored by J. McFadden \cite{mcfadden} in the case of the earliest stage of
biological evolution, where he expresses the revolutionary idea that
the first life appears only in one of innumerous Everett worlds

However, the author dislikes the immediate consequences of his hypothesis which he absolutely correctly deduces: namely, that extraterrestrial life, and therefore, intelligence does not exist (note
the same hypothesis was suggested  for different reasons by I.S. Shklovsky \cite{Shklovsky})
and that life cannot be created in the laboratory

To overcome these obstacles he suggests the another use of quantum theory to explain the improbable event, namely, the inverse Zeno effect. However, we do not consider this opportunity as plausible.

Indeed, he considers as a model of improbable event the passage of light through the
vertically and horizontally polarized lenses while the insertion of extra lenses
between them increase the probability.

This case, however, deals with low-dimensional system when the small probability
is achieved due to a sort of fine-tuning (mutual orthogonality of lenses).
At the same time, the low probability of transition leading to first self-replicator
is due to large dimension of corresponding Hilbert space. More quantitatively,
if one has two wave functions (normalized vectors in a Hilbert space) one of which $|i \rangle$,
corresponds to initial "single amino acid arginine" \cite{mcfadden} while, second, $|f \rangle$,
corresponds to the emerged self-replicator. The typical (average) value of the square of their scalar product, related to a transition probability is
\begin{equation}
< |\langle i | f \rangle|^2 > = \frac{1}{N},
\label{aver}
\end{equation}
where $N$ is a dimension of the Hilbert space defined by the number of participating elements.
Now, if one produce some quantum measurement, the scale of this quantity clearly remains the same.
The only way to increase these probabilities by dense series of measurements would be to
arrange them in some particular way defined by the initial and final states. The appearance of such a special measurement-like process
is not easier to explain than the occurrence of small-probability quantum transition. At the same time, some random measurements will not substantially increase the probability (\ref{aver}), contrary to the case of polarized lenses, when the specially organized low probability may be increased by a generic measurement.

Therefore, we do not consider inverse quantum Zeno effect as a candidate for the explanation of low probability events necessary for life emergence and come back to the initial suggestion of McFadden
about the use of MWI.

Moreover, we suggest to extend this mechanism to all the stages of biological evolution.
Indeed, the original suggestion of \cite{mcfadden} is to limit the field of applicability of
quantum effects to the microbilogical scale \cite{mcfadden1} when the entanglement between
cell and its environment is essential, while for the multi-cell structures quantum effects
were considered \cite{mcfadden} unimportant.

Contrary to that, we suggest that {\it all} the mutations in the course of biological evolution are
the quantum measurent-like processes so that all their different outcomes are realized in different branches.
The increasing of complexity now has purely random character, so that only in few parallel worlds
the biological evolution produces more and more complex species.

All the parallel worlds emerging due to mutation differ only by small variations in the
mutating organism. This feature is common with a standard (neo)Darwinian paradygm.
What is different from it is that {\it all} the versions of this variation are realized in different
parallel Everett worlds. This naturally implies the increase of complexity in some of them just by random process.
In our opinion, this solves the fundamental problem of the extremely law probability of life emergence and
evolution to the mosty complex forms, including ourselves.

There are a number of
fundamental facts which, to our opinion, do not contradict to or even support this hypothesis.
These are "punctuated equilibrium" (evolution proceeds by sudden bursts followed by
long "stasis" periods), "Out of Africa" theory \cite{leakey} (
appearance of all humans from a single family), "Mitochondrial Eve"
(identifying a common female anchestor, being the support of previous theory),
"irreversibility of the brain formation" (once emerged brain never reduced in the course of evolution) etc.

We have no opportunity of detailed discussions and just mention that all these facts
may be understood as emerging from improbable rare events of quantum measurement type,
so that all of their outcomes are realized in parallel worlds. We are just lucky inhabitants of one of the most "pleasant" of them.

\section{Discussion and Conclusions}

In this article we have tried to explore the possible relation between Anthropic Principle and
Many-World Interpretation of Quantum Theory. The key moment is the
possibility to multiply the reality to such an extent that very special events like emergence of Life become quite possible.

The important feature of this process is the smallness of differences between various parallel Everett worlds This allows to scan all the possible values of required parameters which is essentially
similar to the arguments justifying Darwinian natural selection. The only, albeit crucial difference is
that selection occurs not in the different moments of time like Darwinian one, but in
the different parallel worlds, or, mathematically speaking,
in the different regions of Hilbert space.
Such a resemblance to the Darwinian evolution may be explored
for other known mechanisms of generation of variety of options (like string landscape
or eternal chaotic inflation) in order to separate the "physical" predictions from
the effects of "environment" \cite{Rubakov} or "scanning" \cite{Weinberg} which we are
about to suggest.

Indeed, if some physical constant should be fine-tuned for the emergence of life it
is very unlikely that it is completely defined by underlying physics (cf. \cite{Smolin})
and selection process of Darwinian type was likely to contribute.
At the same, the physics should rather lead to the establishing of general framework
and more robust constraints (see, for example, Ref. \cite{Bar-Kam}, where
in the framework of the Euclidean quantum gravity some constraints
on possible values of the effective cosmological constant were found) which
may be a starting point for subsequent fine-tuning by anthropic selection.

In the case of the Many-World Interpretation such a selection allows to fine-tune
various parameters which are not amongst the basic constants of theory of fundamental interactions,
including gravity and elementary particle physics. This is because the branching due to the
Many-Worlds interpretation occurs when all the fundamental constants are already fixed and
therefore they are the same in all the Everett parallel worlds.
We suggested to use the term "Mesoscopic Anthropic Principle" for description of
anthropic selection in the branching process.

We considered two possible fields of applicability of Mesoscopic Anthropic Principle,
namely, planetary coincidences and biological evolution.

In both cases the small differences generated by branching allow to explain the coincidences which is very difficult to do otherwise. As an example  we consider the coincidence of angular sizes of Sun and Moon
responsible for the Solar eclipses. This coincidence may be achieved by small steps during branching,
and anthropic selection may choose it to be realized in our Universe if eclipses played any role
in the life emergence. This hypothesis may be checked , in principle, opening an opportunity for indirect tests
of Anthropic Principle.

The other important problem is the arising of complexity during bilogical evolution,
 including such extreme cases as Life itself and Mind. We suggest that crucial role is played
the Many-Worlds interpretation,
so that extremely small probability is fully compensated by enormous number of trials.

No we are ready to take an hazard to try to give the crudest estimate of number of the Everett worlds
produced up to the present moment. We first assume that it is the Planck constant $\hbar$ which selects the
measurement-like interactions leading to defactorization.
Now, for dimensional reasons when determining the number of worlds
it should be divided by some constant with the dimension of action or phase space, characterizing the whole Universe. The emerging ratio is related to the ratio of the
Planck time $t_P$ and the age of the Universe $T$. Therefore, we expect that the number of worlds $N$
is
\begin{equation}
N = f\left(\frac{T}{t_P}\right).
\end{equation}
where $f$ is some growing function which we allow to range from linear to exponential (the
latter qualitatively supported by the chain character of branching while there is no reason for appearance of
logarithmic function, also growing) which leads to $N$
ranging from $10^{60}$ to $10^{10^{60}}$. Especially the last
number seems to be fairly huge in order to accommodate
all the unlikely events leading to modern picture of Life.

Summing up, we consider the Anthropic Principle combined with the multiple opportunities
opened also by the Many-Worlds interpretation of quantum theory, as new exciting field
of physics and other natural sciences, rather than dull alternative to them.

\section*{Acknowledgment}

O.T. is grateful to Cariplo Science Foundation for support during his stay at the University
of Insubria (Como) and to the Department of Physics and Mathematics of this University for kind hospitality.
This work was partially supported by RFBR Grants N 06-02-16215 and 07-02-91557 and by  LSS-1157.2006.2.

\end{document}